\documentclass[10pt,twocolumn,letterpaper]{article}

\usepackage[final]{cvpr}      

\usepackage{graphicx}
\usepackage{amsmath}
\usepackage{amssymb}
\usepackage{booktabs}
\usepackage{multirow,hhline,array}
\usepackage{kotex}
\usepackage{mathtools}
\usepackage{adjustbox}
\usepackage{makecell}
\usepackage{marvosym}
\usepackage[pagebackref,breaklinks,colorlinks]{hyperref}
\usepackage{enumitem}

\usepackage[capitalize]{cleveref}
\crefname{section}{Sec.}{Secs.}
\Crefname{section}{Section}{Sections}
\Crefname{table}{Table}{Tables}
\crefname{table}{Tab.}{Tabs.}

\usepackage{kotex}

\def\cvprPaperID{6102} 
\def\confName{CVPR}
\def\confYear{2022}

\newcommand{\su}[1]{\textcolor{black}{#1}}

\begin{document}

\title{Local-Selective Feature Distillation for 
Single Image Super-Resolution}

\author{SeongUk Park, Nojun Kwak\\
Seoul National University\\
Seoul, Korea\\
{\tt\small \{swpark0703, nojunk\}@snu.ac.kr}
}
\maketitle

\begin{abstract}
Recent improvements in convolutional neural network (CNN)-based single image super-resolution (SISR) methods rely heavily on fabricating network architectures, rather than finding a suitable training algorithm other than simply minimizing the regression loss. Adapting knowledge distillation (KD) can open a way for bringing further improvement for SISR, and it is also beneficial in terms of model efficiency. KD is a model compression method that improves the performance of Deep Neural Networks (DNNs) without using additional parameters for testing. It is getting the limelight recently for its competence at providing a better capacity-performance tradeoff. In this paper, we propose a novel feature distillation (FD) method which is suitable for SISR. We show the limitations of the existing FitNet-based FD method that it suffers in the SISR task, and propose to modify the existing FD algorithm to focus on local feature information. In addition, we propose a teacher-student-difference-based soft feature attention method that selectively focuses on specific pixel locations to extract feature information. We call our method  local-selective feature distillation (LSFD) and verify that our method outperforms conventional FD methods in SISR problems.
\end{abstract}

\section{Introduction}
\label{sec:intro}

\begin{figure}[t]
\centering
\includegraphics[width=1\linewidth]{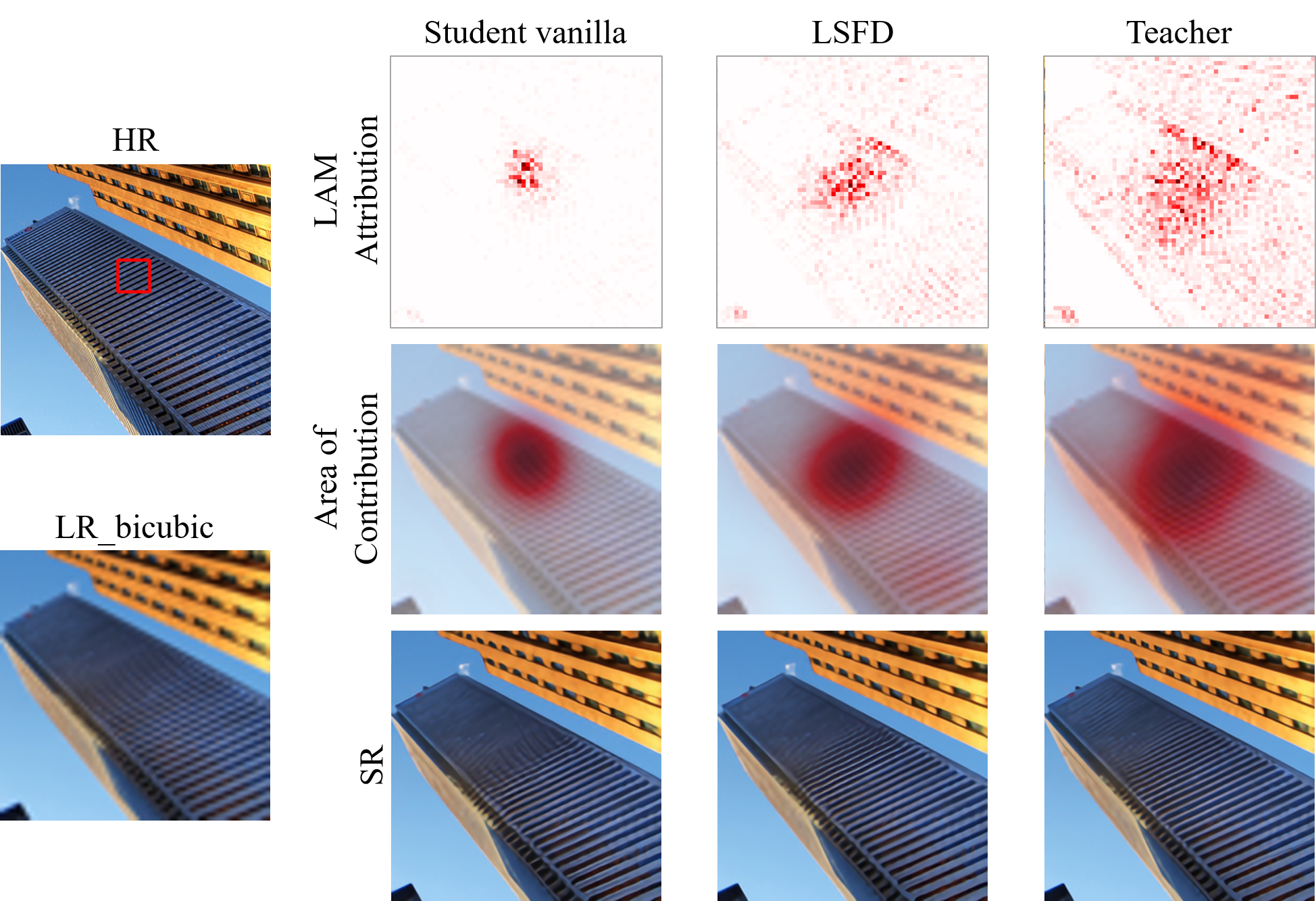}\\
\vspace{-1mm}
\caption{Visualization of {local attribution map} \cite{Gu_2021_CVPR} for the vanilla student network, student network trained using our proposed feature distillation method (LSFD), and teacher network. The student networks are small RCAN \cite{Zhang_2018_ECCV}, and the teacher network is original RCAN, with scale $\times4$ image.}
\label{fig:LAM}
\end{figure}
\vspace{-1mm}
Single image super-resolution (SISR) \cite{freeman2000learning} is an important task in computer vision and image processing, which aims to generate high-resolution images, $I_{SR}$, from degraded low-resolution images, $I_{LR}$, using the ground truth high-resolution images, $I_{HR}$, as the training dataset. 
Among the prior works, SRCNN~\cite{dong2015image} is the first to propose {to use} convolutional neural networks (CNNs) in {SI}SR problems. 
It {outperformed} the other prior works by using only three convolution layers. Afterward, numerous CNN-based super-resolution {(SR)} methods \cite{yang2019deep} that contributed to the progress of super-resolution have been proposed so far. 

EDSR \cite{Lim_2017_CVPR_Workshops} reported that networks with larger capacity and longer training iterations {are} directly related to high performance because deeper super-resolution networks attain larger receptive fields, which enable the network to learn {the} complicated mapping between $I_{LR}$ and $I_{HR}$.
Through residual scaling, EDSR succeeded in stably training the network with deeply stacked residual blocks \cite{he2016deep} without {using} any batch normalization layer, and this brought significant performance improvement. 
After the success of EDSR, networks with a larger number of parameters or computations were preferred for performance improvement.

However, {since the super-resolution techniques are usually used as a preprocessing for a bigger task},
it is also very important whether the trained network is practical with concerning the model size and the computational complexity.
Especially for resource-limited environments such as mobile chips and embedded systems, lightweight networks are generally preferred.
In the field of super-resolution, many studies built lightweight networks to improve model efficiency \cite{dong2016accelerating,Zhang_2018_ECCV,dai2019second,kim2016deeply}.
The studies are generally divided into four categories: 1) efficient network design, 2) model pruning, 3) quantization, and 4) knowledge distillation. 

Among the four methods, we focus on knowledge distillation (KD) \cite{hinton2015distilling}, especially feature-map level knowledge distillation (FD).
Adapting KD for super-resolution is advantageous in two points: 1) Super-resolution generally requires more computations for performance improvements which always have to negotiate with the model efficiency, 2) Introducing additional loss terms that can improve the performance of SISR other than the vanilla $L_1$ loss term has not been well studied.

KD was firstly proposed for the classification problem, and contemporary studies in the field of KD mainly focus on the task of abstraction, such as classification, object detection, and semantic segmentation. To distill knowledge in the tasks other than the classification, feature-map-level knowledge distillation methods are used broadly in order to deliver rich information learned by a stronger teacher network to a student network. 
As we will show later in our ablation studies, the conventional distance-based FD method combined with the regressor, which is composed of $1 \times 1$ convolution layer with ReLU activation \cite{nair2010rectified}, suffers in the task of super-resolution.

As a solution, in this paper, we propose a FD method for a super-resolution model called LSFD, which is an abbreviation of Local-Selective Feature Distillation. 
Existing methods propose mimicking the information at each feature location through a distance loss or distilling information about the relation between every other pixel location for the student network. 
On the other hand, our feature distillation method proposes that information at each pixel location in the feature map only affects the distillation of several neighboring pixels, so that the student network can utilize its receptive fields effectively, which is called local feature distillation (LFD). 
Also, we added an attention function named selective feature distillation (SFD) based on the teacher-student error to further enhance our feature distillation performance. By binding them, we propose a novel feature distillation method designed for SISR, called local-selective feature distillation (LSFD). 
An example of visualizing our method is in Figure \ref{fig:LAM}, using the method of LAM \cite{Gu_2021_CVPR}, showing that our method allows the student model to well-utilize its receptive fields. We will discuss more and show more examples in supplementary materials. The main contributions of our paper are:
\begin{itemize}[leftmargin = *]
    \item We propose local feature distillation (LFD), which is a straightforward method that expands the receptive field of feature distillation locally to alleviate the problems of the conventional distance-based feature distillation methods in super-resolution.
    \item We also propose selective feature distillation (SFD) that uses an attention mechanism for the feature maps in the knowledge distillation, which grants further improvements to LFD, called local-selective feature distillation (LSFD).
    \item We demonstrate the effectiveness of our LSFD on several well-known benchmark models and datasets, and achieve performance improvements at single image super-resolution on most benchmarks having different resolutions.
\end{itemize}

\section{Related Work}
\subsection{CNN-based Super-resolution}\label{sec:2_1}
CNN-based models learn the mapping from $I_{LR}$ to $I_{HR}$ directly. After SRCNN \cite{dong2015image} which used a simple CNN model for SR, many deep-learning-based algorithms outperformed traditional methods \cite{yang2008image,zhang2006edge,huang2015single} by far, and many CNN-based algorithms \cite{Lim_2017_CVPR_Workshops,Zhang_2018_ECCV,dai2019second,kim2016accurate,kim2016deeply,haris2018deep,zhang2018residual,tai2017memnet} contributed to performance improvement in the field of SISR. 
Though SRCNN expected that stacking more layers would lead to performance improvement, this has not been easily achieved due to the vanishing gradient problem. Later, this problem was solved due to the advent of residual blocks \cite{he2016deep}. 
Using residual blocks, VDSR \cite{kim2016accurate}, and DRCN \cite{kim2016deeply} developed networks with deeply stacked residual blocks to improve performance. 
Furthermore, EDSR \cite{Lim_2017_CVPR_Workshops} largely improved the performance of SISR by eliminating the batch normalization \cite{ioffe2015batch}, claiming that the batch normalization regularizes the representation power of the model. 
However, it was difficult to stably train a deep model without batch normalization. 
For a solution, EDSR suggested using residual scaling, and succeed in stably training a deep and wide network with 32 residual blocks and 256 channels. 

Relatively recently, methods adopting various attention mechanisms \cite{bahdanau2014neural,xu2015show} are thriving in the field of single image super-resolution. 
RCAN \cite{Zhang_2018_ECCV} used residual channel attention blocks and showed impressive performance enhancement. Though few years have passed, RCAN still serves as a strong baseline method that frequently takes the second-best place in many state-of-the-art benchmarks \cite{dai2019second,niu2020single}. 
RAM \cite{kim2018ram} used intra-channel attention as well as inter-channel attention and recorded comparable performance. Algorithms using spatial attention have also been proposed. SAN \cite{dai2019second} proposed second-order channel attention (SOCA) and non-locally enhanced residual group (NLRG) to refine features using feature statistics.
HAN \cite{niu2020single} proposed {the} layer attention module (LAM) and {the} channel-spatial attention module (CSAM) that collaboratively {consider} multi-scale layers, and improved the SISR results. However, spatial attention{s} or non-local operations {come} with many shortcomings: they occupy huge memory and slow down the inference speed in GPUs at test time, which {limit their usage} in practice.

\subsection{Effieicnt Single Image Super-resolution}
Given a low-resolution image $I_{LR}$, a SISR network outputs a super-resolution image $I_{SR}$ that aims to reconstruct the high-resolution image $I_{HR}$. 
In the training of a network for SISR, it is common to use cropped patches of the original images as $I_{LR}$ and $I_{HR}$, and for testing, the network uses the whole original image as $I_{LR}$. 
Since the SISR network only learns with the cropped small patches, there is a large discrepancy between the training and testing. 
A common phenomenon is that the network fails to reconstruct the local patterns at the test phase, even though it can be inferred from the neighboring pixels in a human sense. Thus, many recent works in SISR \cite{dai2019second,niu2020single,liu2018non,Mei_2021_CVPR} proposed using non-local operations to enhance the performance.
However, non-local operations involve extra operations that often require huge computations at test time since the complexity of a non-local operation may increase at least asymptotically linear \cite{Mei_2021_CVPR} or quadratic \cite{dai2019second} to the input size, i.e. $O(NC)$ or $O(N^2C^2)$, where $N = H \times W$ when the height of the input image is $H$, and the width is $W$, and $C$ is the channel size of the SISR network. 
Although the non-local operation may not largely increase the parameter of the network, forwarding large test images occupies huge memory and requires lots of time consumption. 
Thus, networks with good computation-to-performance tradeoffs are preferred in most situations but will be particularly preferred in resource-limited embedded systems, and many lightweight networks are recently proposed \cite{liu2021splitsr,chen2021attention,guo2020hierarchical,luo2020latticenet} that focus on designing a good network for efficient super-resolution. 

\subsection{Feature-map-level Knowledge Distillation}\label{SEC:KD}
KD \cite{hinton2015distilling} proposed training a student network to learn soft-labels extracted from a larger teacher network, and achieved meaningful improvements. 
One shortcoming of this label-based approach is that the original KL-divergence-based loss is not applicable to tasks other than image classification, so that feature-map-level distillation would be preferred for researchers who want to apply knowledge distillation to other forms of tasks. 
In the early studies, AT \cite{zagoruyko2016paying} proposed to distill simple attention map, which is made by simply summing up the channel activations in the channel direction, and achieved comparable results to KD. 
FitNet \cite{romero2014fitnets} proposed using both label information and feature information. 
For feature information, they proposed {to use a} regressor, which is made of a $1 \times 1$ convolution layer, claiming that it can cope with the channel difference between the teacher network and the student network. 
This simple transformation using $1 \times 1$ convolution became the key element in future feature distillation researches \cite{kim2018paraphrasing,heo2019comprehensive,park2020feature,chung2020feature,heo2019knowledge}.
This regressor is also used in other studies that adapted feature distillation, such as object detection \cite{chen2017learning,Wang_2019_CVPR,Dai_2021_CVPR,zhang2021improve}. 

For super-resolution, Zhang et al. \cite{zhang2021data} proposed distilling knowledge in super-resolution with training data. 
Lee et al. \cite{lee2020learning} enhanced FSRCNN \cite{dong2016accelerating} using an autoencoder \cite{masci2011stacked} that is trained by reconstructing the $I_{HR}$, and trained the FSRCNN to mimic the intermediate feature of the autoencoder. 
The most related study to our work is FAKD \cite{he2020FAKD}, which proposed distilling the feature-affinity matrix of the teacher network to the student network, which improved the performance of SAN and RCAN.

\section{Proposed Method}
In this work, we focus on a promising solution that has not been spotlighted much in SISR: knowledge distillation which makes the trained network more efficient, and operates in the level of the loss function. To the best of our knowledge, only a few studies \cite{lee2020learning,he2020FAKD} exist that adopt knowledge distillation for super-resolution. 

\subsection{Knowledge distillation for SISR}
Adapting knowledge distillation in the field of single image super-resolution is promising in two aspects. \textbf{First}, whereas finding an efficient network architecture for SISR has been extensively studied, researches that seek for finding improved training methodologies other than a simple $L_1$ regression loss are lacking.
Since knowledge distillation generally grants performance improvements with additional loss terms to the conventional loss term, adopting knowledge distillation can bring further improvements in the field of single image super-resolution.
\textbf{Second}, using knowledge distillation for SISR is meaningful for the model in the view of efficiency. 
As explained in Section \ref{sec:2_1}, recent developments in SISR mostly depend on non-local operations for their performance improvements \cite{dai2019second,niu2020single,kim2018ram,Mei_2021_CVPR,liu2018non}. 
Although non-local operations are parameter-efficient because non-local operations do not use additional parameters, they occupy huge GPU memory at test time, which will be shown in Section \ref{sec:exp_set}. 

To adapt knowledge distillation, a straightforward approach is using the output super-resolution image of the teacher $I_{SR}^{T}$ as the learning target. Thus the loss function for the student $I^{S}_{SR}$ is: 
{\small
\begin{equation}
    \mathcal{L}_{SR}= \frac{1}{2N} ( \sum_{n=1}^{N} \| {I^{T}_{SR}}_n - {I^{S}_{SR}}_n\|_{1} +
    \sum_{n=1}^{N} \| {I_{HR}}_n - {I^{S}_{SR}}_n\|_{1} ),
\end{equation}
\label{eq:LFD}
}
where $N$ refers to the number of training samples in a mini-batch.
It is also worth considering using rich feature information, called feature distillation.
One plausible and straightforward option to consider when adapting feature distillation is to use the regressor \cite{romero2014fitnets} which was mentioned in Section \ref{SEC:KD}, because it has shown its effectiveness on the tasks other than image classification \cite{chen2017learning,Wang_2019_CVPR,Dai_2021_CVPR,zhang2021improve}. 
However, it has shown to be ineffective in the experiment {in} FAKD \cite{he2020FAKD}. 
We also report this problem later in our experiment.

\begin{figure*}[t]
\centering
\includegraphics[width=1\linewidth]{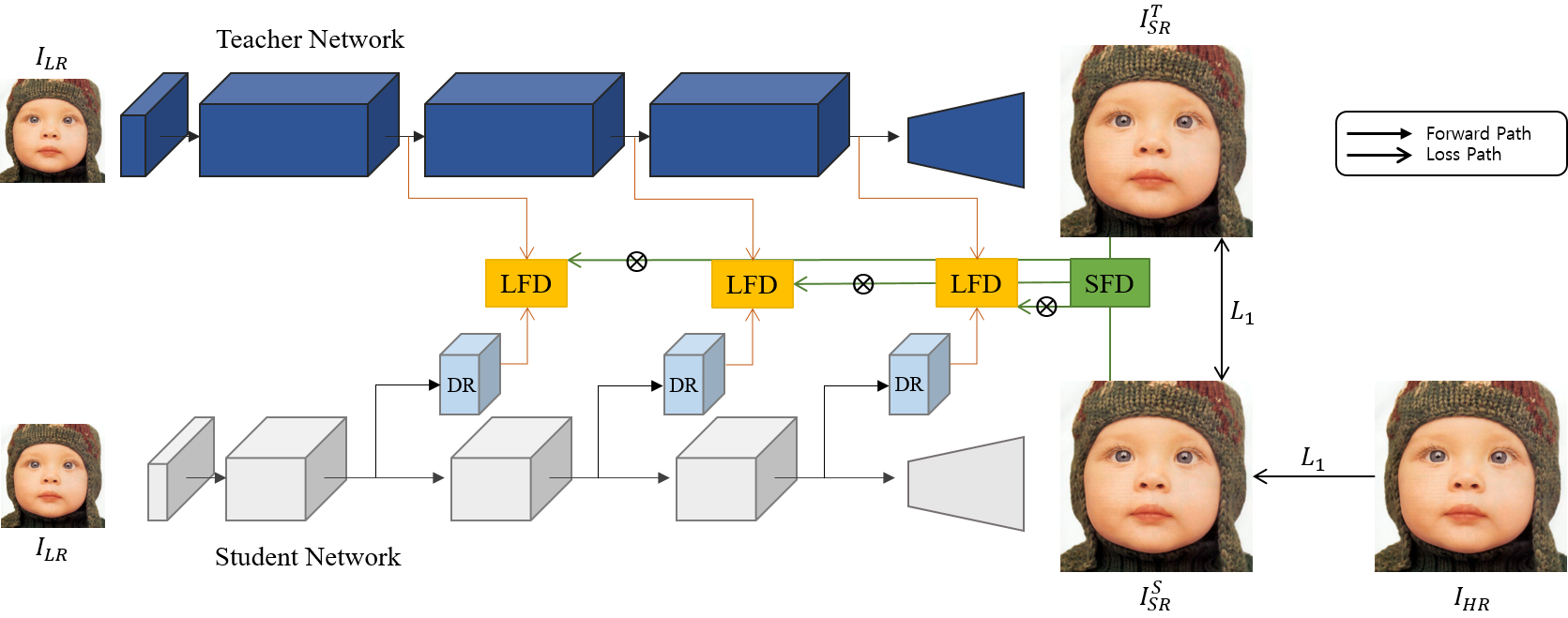}\\
\vspace{-1.5mm}
\caption{Overall illustration of the training of LSFD. LSFD consists of LFD + SFD. The detailed structure of SFD is depicted in \cref{fig:SFD}.}
\label{fig:LSFD}
\end{figure*}
\vspace{-1.5mm}

\subsection{Local Feature Distillation}
\label{sec:LFD}
We conjecture that the reason FitNet \cite{romero2014fitnets}, which worked well in the classification tasks and became the basis of many feature distillation papers \cite{kim2018paraphrasing,heo2019comprehensive,park2020feature,chung2020feature,heo2019knowledge,chen2017learning,Wang_2019_CVPR,Dai_2021_CVPR,zhang2021improve}, suffers in super-resolution is owing to the limited receptive field. From the receptive field point of view, classification has two major differences from super-resolution. 

First, networks that record high performance in classification tasks are deep enough that the receptive field does not have to be considered. On the other hand, in super-resolution, the batch-normalization layer, a form of a regularizer, is not used due to the performance issue. Henceforth, it is difficult to stably train a very deep network. However, relatively shallow networks incur the problem of a limited receptive field.

Second, in image classification, the test image is resized to be the same size as the learned image, wherever in single image super-resolution, an $I_{LR}$ typically cropped by $48 \times 48$  is used for the network training, and a large full $I_{LR}$ is used for testing. 
In conventional single image super-resolution, this discrepancy between training and testing has been tolerated for the reason that CNNs are translation invariant. 
However, it is problematic from the receptive field point of view. 

In the network training of SISR, the network is not forced to effectively utilize the receptive fields because only small-sized $I_{LR}$ are used for training, while effectively utilizing larger receptive fields will be essential for large test images.
Furthermore, if we want to distill knowledge for model compression, the student network is more likely to be a shallower network, exacerbating this receptive field problem.

To alleviate this receptive field problem, we propose a simple modification to the regressor of the FitNet: we use a deeper regressor for the training of the student network. 
Since the student features at each feature map location are involved {in} mimicking larger feature map area of the teacher network, they are forced to utilize their limited receptive fields effectively. 
Our method is inspired by the work of Gu et al. \cite{Gu_2021_CVPR}, which visualized that compared to the traditional deep super-resolution models such as SRCNN and EDSR, recent non-local super-resolution networks such as RNAN\cite{zhang2019residual} and SAN utilize a much wider range of information.
For each particular position in the feature map of the student network, more comprehensive feature map positions surrounding the corresponding pixel position in the feature map of the teacher network are engaged in the feature distillation. 
We call this method \textit{Local Feature Distillation} (LFD), and its distillation loss function becomes:
\begin{equation}
\begin{split}
& D = \frac{F_T}{\|F_T\|_2}-\frac{DR(F_S)}{\|DR(F_S)\|_2}, \\
& \mathcal{L}_{LFD}=\alpha_{1} \:  \| D \|_1
\end{split}
\label{eq:LFD}
\end{equation}
where the $F_T$ and $F_S$ refer to the feature map of the teacher and student network, respectively, and $DR(\cdot)$ refers to the deeper regressor which consists of five $3 \times 3$ convolution layers, each having a Leaky-ReLU activation with the slope of 0.1. The feature maps are normalized for better distillation as proposed in AT \cite{zagoruyko2016paying}. $\alpha_{1}$ is a hyper-parameter for scaling the loss term. 
It is called LFD because \su{consecutive} $3\times3$ convolutions gather local information of the student network.
The illustration of our overall training is shown in Figure \ref{fig:LSFD}. 

Distilling local features is also very intuitive in the human sense. When $I_{LR}$ whose local repetitive pattern has been blurred is to be restored to $I_{SR}$, the information about the erased pattern could be inferred using the information in the surrounding pattern, especially when there exists a specific pattern that repeats within an image.



\subsection{Selective Feature Distillation}

\begin{figure}[t]
\centering
\includegraphics[width=1\linewidth]{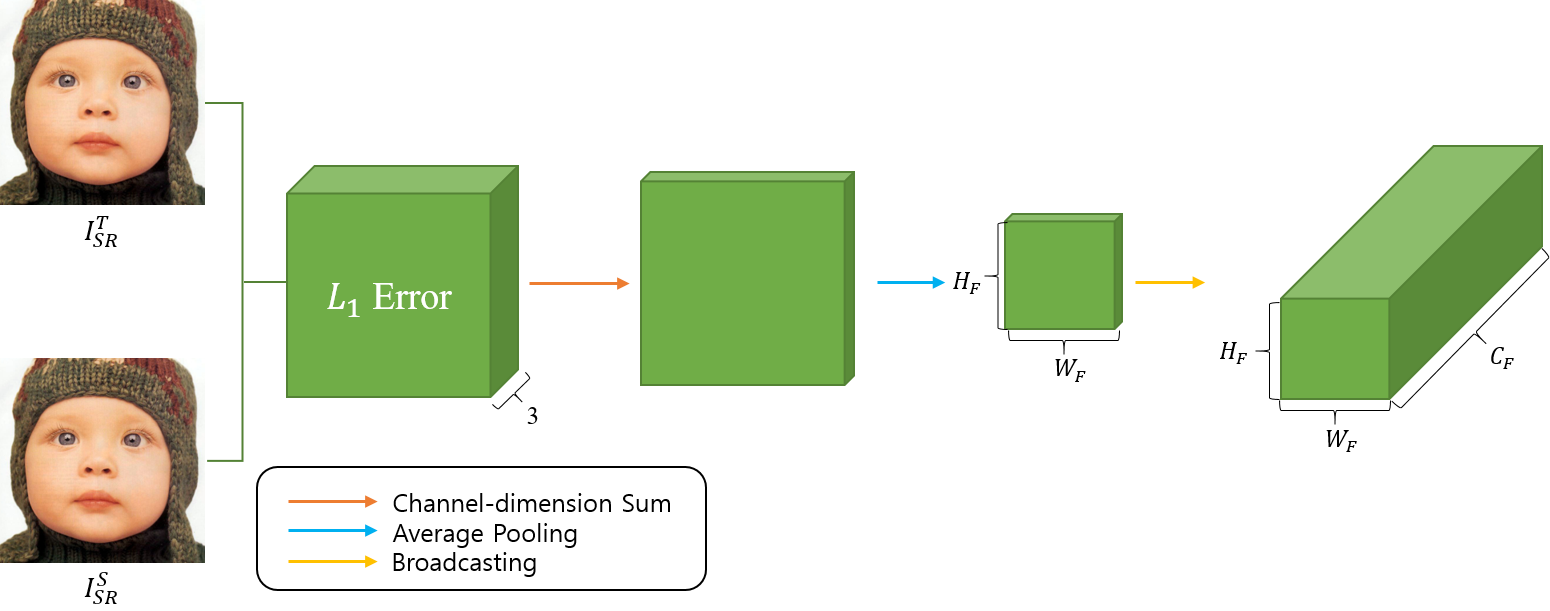}\\
\vspace{-1.0mm}
\caption{Detailed illustration of SFD. $H_{F}$, $W_{F}$, and $C_{F}$ refers to the height, width, and the channel size of the distilled feature maps at the  intermediate layers.}
\label{fig:SFD}
\end{figure}
\vspace{-1.5mm}
To further improve LFD, we propose {\textit{selective feature distillation} (SFD) which calculates} an attention map for the feature distillation location. Merging LFD and SFD together, we call our method \textit{local-selective feature distillation} (LSFD). 
Because FD adds an auxiliary loss to the conventional \su{$\mathcal{L}_{SR}$} loss, it is natural for FD to attend more to the feature map positions where the student and the teacher disagree much and let the \su{$\mathcal{L}_{SR}$} loss take care of the other positions.    
This motivates us to pay more attention to the positions where the difference between $I^{T}_{SR}$ and $I^{S}_{SR}$ is large.
For SFD, the detailed process for obtaining the attention map is {depicted in} Figure \ref{fig:SFD}, and it is formulated as: 
\begin{equation}
    SFD = \text{B}(\text{P}(\sum_{n=1}^{3}{\| {I^{T}_{SR}}_n - {I^{S}_{SR}}_n\|_{1}}, S), C),
\end{equation}
\label{eq:SFD}
where $\text{P}(\cdot, S)$ is spatial average pooling over the tensor with a kernel size of $S$, and $S$ refers to the scale ratio of the model training, and $\text{B}(\cdot), C$ {represents the} broadcasting operation over the channel dimension with $C$ {being} the channel size of the feature map in the SISR network. Since SISR networks do not increase the spatial dimension of the intermediate features and only adapt the pixelshuffle operation at the end of the network, the pooled and broadcasted tensor has the same size as the intermediate feature maps.

This is a big difference compared to FAKD \cite{he2020FAKD}, which proposed to learn the feature correlation of different feature map positions by distilling a correlation matrix with $HW \times HW$ size. 
In FAKD, calculating correlation matrix is a non-local operation, so that the student network is forced to learn feature relations from every feature position pair, even though there may exist unnecessary pairs that are less related, i.e. features on different positions that are distant, or features that are not lying on the same object. 
We thought that it would be more helpful to focus on the more informative pixels. To this end, we use SFD as an attention operator to {LFD presented in \cref{sec:LFD}.}

The final LSFD loss is derived by element-wise multiplication of $SFD$ and $D$ matrix. Combining all the loss terms together, the final total loss is:
\begin{equation}
\mathcal{L}_{LSFD}= \alpha_{2} \: \| SFD \otimes D \|_{1}\\ 
\label{eq:LSFD}
\end{equation}
\begin{equation}
\mathcal{L}_{total}= \mathcal{L}_{LSFD} + \mathcal{L}_{SR} 
\label{eq:total}
\end{equation}
where $\otimes$ denotes to the element-wise multiplication, and $\alpha_{2}$ is a hyper-parameter for scaling the $\mathcal{L}_{LSFD}$ to match the loss scale of $\mathcal{L}_{SR}$.

\section{Experiments}
In this section, we first explain the details of our network training, and analyze the experimental results both quantitatively and qualitatively. 

\subsection{Experiment Settings}\label{sec:exp_set}
 \begin{table}[t]
\begin{center}
\resizebox{1.0\linewidth}{!}{
\begin{tabular}{|c|c|c|c|c|c|c|}
\Xhline{3\arrayrulewidth}
\multirow{2}*{Model Configuration} & \multicolumn{2}{c|}{RCAN}    & \multicolumn{2}{c|}{EDSR}   & \multicolumn{2}{c|}{SAN}  \\
\cline{2-7}
               & T         & S          & T          & S              & T          & S      \\
\Xhline{3\arrayrulewidth}
Channel Size   & 64        & 64         &256         & 64             & 64         & 64\\
Resblocks      & 20        & 6          & 32         & 16             & 10         & 10\\
ResGroups      & 10        & 10         & -          & -              & 20         & 6\\
\Xhline{1\arrayrulewidth}
Params (M)     & 15.59     & 5.17       & 43         & 1.5            & 15.86      & 5.0          \\
Runtime (ms)   & 117       & 42         & 171        & 14             & 2670       & 806          \\
GPU Memory (GiB)   & 1.29      & 1.20       & 1.83       & 1.18           & 17.75      & 17.71        \\
MultiAdds (T)  & 4.70      & 1.53       & 12.51      & 0.42           & -          & -            \\
\Xhline{3\arrayrulewidth}
\end{tabular}
}
\end{center}
\vspace{-1.5mm}
\caption{Model configurations and their efficiency indicators for the RCAN, EDSR, and SAN networks that were experimented {in} this paper. }
\label{table:Eff}
\end{table}

Following the previous literature \cite{dong2015image,kim2016accurate,Lim_2017_CVPR_Workshops,Zhang_2018_ECCV,dai2019second}, we use 800 images from DIV2K \cite{timofte2017ntire} dataset for the network training, and use Set5, Set14, BSD100 and Urban100 dataset for testing the trained network, each having different characteristics. 
We conducted experiments on three SISR networks, RCAN \cite{Zhang_2018_ECCV}, EDSR \cite{Lim_2017_CVPR_Workshops}, and SAN \cite{dai2019second}. 
All the student networks are trained using ADAM \cite{kingma2014adam} optimizer with $\beta_{1} = 0.9$, $\beta_{2} = 0.99$, and $\epsilon = 10^{-8}$. 
The initial learning rate is 0.0001, and is halved at 150 epochs. The main comparator in our experiments was FAKD \cite{he2020FAKD}. 
FAKD trained the student network for 200 epochs, but empirically, this was far from the point of saturation. 
Therefore, we trained all the student networks for 300 epochs. 
All the networks use minibatches containing 16 low-resolution patches whose size is $48\times48$ for input, and only horizontal flip was applied for data augmentation. 
Random rotation was not applied to exclude possible interactions with feature distillation, which is different from the training of the original SAN. 
For LFD, $\alpha_1$ is set to 2,000, and $\alpha_2$ is set to 10 in  \cref{eq:LFD} and \cref{eq:LSFD}.

For the network architecture, the teacher network of the RCAN has 20 residual channel attention blocks (RCAB) in every 10 residual groups (RG) and 64 feature map channels. For the student network of RCAN, the number of RCABs in each RG is changed to 6; thus, the student network has about 1/3 parameters compared to the teacher network. In contrast, for SAN, we maintain the residual blocks in each local-source residual attention group (LSRAGs) as 10 but reduce the number of LSRAGs from 20 for the teacher network to 6 for the student network. For both types of networks, we maintain the number of channels for the student networks. For EDSR, we reduce the number of residual blocks and the number of channels from 32 and 256 to 16 and 64. 
 
Some measures for the models which we experimented with are provided in Table \ref{table:Eff}, together with the model configurations explained above.
The `Memory' refers to the amount of GPU memory consumption at the test phase, and the runtime is the average time taken for processing an input image. 
The runtime and memory usage are measured using 100 images in the B100 benchmark dataset with the upscale ratio of 2 and batch size of 1, using a single Geforce Titan RTX GPU. 
The MultiAdds is calculated using single $3\times640\times480$ sized input, and the GPU memory consumption in the table is the ceiling value required to measure the entire data set. 
Since EDSR only has resblocks, numbers on the ResGroups row are omitted. 
For SAN, the SOCA block of SAN has intractable calculations because it handles heavy operations, which includes ${HWC}\times{HWC}$ sized covariance matrix, so the MultiAdds row is omitted.

\subsection{Quantitative Results}
\begin{table}[h]
\begin{center}
\resizebox{1.0\linewidth}{!}{
\begin{tabular}{|c|c|c|c|c|c|}
\Xhline{3\arrayrulewidth}
\multicolumn{6}{|c|}{RCAN}\\
\Xhline{1\arrayrulewidth}
Scale                      & Methods    & Set5   & Set14  & B100   & Urban100  \\
\Xhline{3\arrayrulewidth}
\multirow{9}*{$\times$2}    & Teacher    & 38.271 & 34.126 & 32.390 & 33.176\\
\cline{2-6}
                            & Student    & 38.074 & 33.623 & 32.199 & 32.317\\
                            & FAKD       & 38.164 & 33.815 & 32.274 & 32.533\\
                            & FAKD*      & \textcolor{blue}{38.180} & 33.828 & 32.284 & 32.602\\
                            & FitNet     & 38.132 & 33.759 & 32.253 & 32.460\\
                            & LFD        & 38.178 & \textcolor{blue}{33.840} & \textcolor{red}{32.296} & \textcolor{blue}{32.669}\\
                            & LSFD       & \textcolor{red}{38.189} & \textcolor{red}{33.882} & \textcolor{blue}{32.291} & \textcolor{red}{32.704}\\
                            \cline{2-6}
                            & FAKD + FFT & 38.187 & 33.866 & 32.289 & 32.669\\
                            & LSFD + FFT & 38.181 & 33.847 & \textbf{32.298} & 32.667\\
\Xhline{3\arrayrulewidth}
\multirow{9}*{$\times$3}    & Teacher    & 34.758 & 30.627 & 29.309 & 29.104 \\
\cline{2-6}
                            & Student    & 34.557 & 30.408 & 29.162 & 28.482 \\
                            & FAKD       & 34.653 & 30.449 & 29.208 & 28.523 \\
                            & FAKD*      & \textcolor{red}{34.667} & 30.490 & 29.209 & 28.611\\
                            & FitNet     & 34.570 & 30.466 & 29.184 & 28.493\\
                            & LFD        & 34.657 & \textcolor{red}{30.525} & \textcolor{blue}{29.224} & \textcolor{blue}{28.665}\\
                            & LSFD       & \textcolor{blue}{34.666} & \textcolor{blue}{30.510} & \textcolor{red}{29.226} & \textcolor{red}{28.689} \\
                            \cline{2-6}
                            & FAKD + FFT & 34.666 & \textbf{30.527} & 29.218 & 28.635 \\
                            & LSFD + FFT & 34.658 & 30.508 & \textbf{29.227} & 28.666 \\
\Xhline{3\arrayrulewidth}
\multirow{9}*{$\times$4}    & Teacher    & 32.638 & 28.851 & 27.748 & 26.748 \\
\cline{2-6}
                            & Student    & 32.321 & 28.688 & 27.634 & 26.340 \\
                            & FAKD       & 32.462 & 28.750 & 27.678 & 26.422 \\
                            & FAKD*      & 32.461 & \textcolor{blue}{28.779} & 27.685 & 26.490\\
                            & FitNet     & 32.417 & 28.716 & 27.660 & 26.406\\
                            & LFD        & \textcolor{blue}{32.475} & \textcolor{red}{28.783} & \textcolor{blue}{27.693} & \textcolor{red}{26.542}\\
                            & LSFD       & \textcolor{red}{32.497} & 28.771 & \textcolor{red}{27.699} & \textcolor{blue}{26.525} \\
                            \cline{2-6}
                            & FAKD + FFT & 32.491 & 28.755 & 27.693 & 26.507\\
                            & LSFD + FFT & \textbf{32.513} & 28.774 & \textbf{27.709} & 26.535 \\
 \Xhline{3\arrayrulewidth}
\end{tabular}
}
\end{center}
\vspace{-1.5mm}
\caption{PSNR scores measured on RCAN networks. The scores on the FAKD row are from its original paper, and the scores on FAKD* row are our reproduced scores with longer training epochs. The red and blue text is the best and second-best scores in different combinations of datasets and scales. For the models trained using FFT loss, the scores that outperformed the best scores among the other models without FFT were highlighted in bold fonts.}
\label{table:RCAN}
\end{table}

The quantitative results are in Table \ref{table:RCAN}, \ref{table:EDSR}, \ref{table:SAN}. 
In this subsection, we analyze the results on these tables based on several different criteria.

\begin{table}[h]
\begin{center}
\resizebox{1.0\linewidth}{!}{
\begin{tabular}{|c|c|c|c|c|c|}
\Xhline{3\arrayrulewidth}
\multicolumn{6}{|c|}{EDSR}\\
\Xhline{1\arrayrulewidth}
Scale                      & Methods    & Set5   & Set14  & B100   & Urban100  \\
\Xhline{3\arrayrulewidth}
\multirow{7}*{$\times$2}    & Teacher    & 38.090 & 33.797 & 32.241 & 32.373\\
\cline{2-6}
                            & Student    & 37.919 & 33.478 & 32.126 & 31.840\\
                            & FAKD*       & 37.976 & 33.523 & \textcolor{blue}{32.156} & \textcolor{blue}{31.906}\\
                            & LFD        & \textcolor{blue}{37.984} & \textcolor{red}{33.547} & \textcolor{blue}{32.156} & 31.896\\
                            & LSFD       & \textcolor{red}{37.991} & \textcolor{blue}{33.529} & \textcolor{red}{32.157} & \textcolor{red}{31.936}\\
                            \cline{2-6}
                            & FAKD + FFT & \textbf{38.004} & {23.547} & {32.156} & {31.928}\\
                            & LSFD + FFT & 37.983 & \textbf{33.555} & \textbf{32.161} & \textbf{31.976}\\
\Xhline{3\arrayrulewidth}
\multirow{7}*{$\times$3}    & Teacher    & 34.547 & 30.435 & 29.167 & 28.470 \\
\cline{2-6}
                            & Student    & 34.272 & 30.266 & 29.044 & 27.959 \\
                            & FAKD*       & \textcolor{blue}{34.356} & \textcolor{blue}{30.296} & 29.066 & \textcolor{blue}{28.016} \\
                            & LFD        & 34.348 & 30.287 & \textcolor{blue}{29.068} & 27.999 \\
                            & LSFD       & \textcolor{red}{34.384} & \textcolor{red}{30.302} & \textcolor{red}{29.077} & \textcolor{red}{28.029} \\
                            \cline{2-6}
                            & FAKD + FFT & {34.356} & \textbf{30.312} & \textbf{29.078} & \textbf{28.057} \\
                            & LSFD + FFT & \textbf{34.382} & \textbf{30.313} & \textbf{29.084} & \textbf{28.080} \\
\Xhline{3\arrayrulewidth}
\multirow{7}*{$\times$4}    & Teacher    & 32.385 & 28.741 & 27.661 & 26.425 \\
\cline{2-6}
                            & Student    & 32.102 & 28.526 & 27.538 & 25.905 \\
                            & FAKD*       & \textcolor{red}{32.138} & \textcolor{blue}{28.547} & \textcolor{blue}{27.557} & \textcolor{blue}{25.972} \\
                            & LFD        & \textcolor{blue}{32.107} & 28.524 & 27.552 & 25.962\\
                            & LSFD       & \textcolor{blue}{32.107} & \textcolor{red}{28.548} & \textcolor{red}{27.563} & \textcolor{red}{25.980} \\
                            \cline{2-6}
                            & FAKD + FFT & \textbf{32.174} & \textbf{28.556} & {27.562} & \textbf{26.008}\\
                            & LSFD + FFT & \textbf{32.140} & \textbf{28.561} & \textbf{27.570} & \textbf{26.003} \\
 \Xhline{3\arrayrulewidth}
\end{tabular}
}
\end{center}
\vspace{-1.5mm}
\caption{PSNR scores measured on EDSR networks. The meaning of the asterisk, blue, red, bold texts are the same as Table \ref{table:RCAN}.
}
\label{table:EDSR}
\end{table}

\begin{table}[h]
\begin{center}
\resizebox{1.0\linewidth}{!}{
\begin{tabular}{|c|c|c|c|c|c|}
\Xhline{3\arrayrulewidth}
\multicolumn{6}{|c|}{SAN}\\
\Xhline{1\arrayrulewidth}
Scale                       & Methods    & Set5   & Set14  & B100   & Urban100  \\
\Xhline{3\arrayrulewidth}
\multirow{4}*{$\times$2}    & Teacher    & 38.144\Cross & 33.806\Cross & 32.378 & 32.673\Cross \\
                            & Student    & 38.039\Cross & 33.700\Cross & 32.211 & 32.358\Cross\\
                            & FAKD*       & 38.107\Cross & 33.738\Cross & 32.267 & 32.558\Cross\\
                            & LSFD       & \textcolor{red}{38.131}\Cross & \textcolor{red}{33.781}\Cross & \textcolor{red}{33.276} & \textcolor{red}{32.562}\Cross\\
\Xhline{3\arrayrulewidth}
\multirow{4}*{$\times$3}    & Teacher    & 34.642 & 30.492 & 29.206 & 29.166\Cross \\
                            & Student    & 34.461 & 30.397 & 29.129 & 28.394\Cross \\
                            & FAKD*       & \textcolor{red}{34.638} & 30.464 & 29.194 & 28.540\Cross \\
                            & LSFD       & 34.627 & \textcolor{red}{30.468} & \textcolor{red}{29.197} & \textcolor{red}{28.594}\Cross \\
\Xhline{3\arrayrulewidth}
\multirow{4}*{$\times$4}    & Teacher    & 32.398 & 28.754 & 27.696 & 26.557\Cross \\
                            & Student    & 32.298 & 28.683 & 27.622 & 26.329\Cross \\
                            & FAKD*       & 32.380 & \textcolor{red}{28.746} & 27.679 & \textcolor{red}{26.466}\Cross \\
                            & LSFD       & \textcolor{red}{32.413} & 28.733 & \textcolor{red}{27.681} & 26.453\Cross \\
 \Xhline{3\arrayrulewidth}
\end{tabular}
}
\end{center}
\vspace{-1.5mm}
\caption{PSNR scores measured on SAN networks. The numbers with \Cross means it was tested in four-crops, and the FAKD* is our reproduced result.}
\label{table:SAN}
\end{table}

\noindent\textbf{Impact of deeper regressor:} From Table \ref{table:RCAN}, we can compare the result of FitNet and LFD. When FitNet is used directly on SISR, a slight increase in the PSNR scores occurs. By making the regressor much deeper, each pixel learns from more spacious features from the teacher network so that the student network learns to focus on local regions that may help reconstruct local patterns. Thus, LFD is much more suitable for SISR, leading to the success of feature distillation in super-resolution.

\noindent\textbf{Impact of LFD and LSFD:}
In Table \ref{table:RCAN} and \ref{table:EDSR}, the test PSNR estimated using our LSFD outperforms the previous methods in most of the experiments. Our LFD and LSFD especially perform better in Urban100 dataset, where reconstructing repetitive local patterns is essential, as we intended. However, this effect diminishes in a high scale ratio ($\times4$). We conjecture that the information loss is so significant that the local patterns are no longer helpful in reconstructing the original. 
For all training methods, the increases in PSNR in EDSR are poorer on average compared to RCAN, but LSFD performed better than others in most settings. We conjecture that the reason for the inferior feature distillation performances in EDSR is using a long skip connection instead of several short skip connections. The LFD seems to be inferior compared to FAKD in EDSR models, but by introducing SFD, LSFD overtakes FAKD.

\noindent\textbf{Adding Frequency-domain loss:}
We also tried using the frequency domain loss and reported the results in Table \ref{table:RCAN} and \ref{table:EDSR}. 
There were a few papers that proposed using frequency domain information. Li et al. \cite{li2018frequency} proposed using frequency domain inputs for super-resolution, and some other papers \cite{sims2020frequency,fuoli2021fourier} suggested frequency component of the $I_{SR}$ for perceptual loss in perceptual super-resolution. In our experiments, we simply use Fourier transformation for our loss:
\begin{equation}
    \mathcal{L}_{FFT} = \|\text{FFT}(I^{S}_{SR}) - \text{FFT}(I^{T}_{SR})\|_{1},
\end{equation}
where $\text{FFT}(\cdot)$ is fast Fourier transformation.
Using $\mathcal{L}_{FFT}$ was not much help for the RCAN models but was effective for EDSR models. The usage of $\mathcal{L}_{FFT}$ had desirable synergy with both LSFD and FAKD.

\noindent\textbf{Results on SAN:}
For the results in Table \ref{table:SAN}, there exist limits for a fair comparison because some results contain 4-crop testing, which means we divide each test {image} in quarter size to generate four of small $I_{SR}$s, and attached them together later.
This drops PSNR because the network cannot get full advantage of non-local operation. 
However, it was inevitable in order to measure the PSNR because SAN networks require a huge number of operations with $O(N^{2}C^{2})$ complexity, and test images often have a high spatial dimension. 
For the numbers with * marks, PSNR was measured by dividing $I_{LR}$ into quadrants and stitching four small SRs since it was impossible to forward full $I_{LR}$ image using a minibatch of size 1 with RTX A6000 GPU which has 48GB memory.

\subsection{Qualitative Results}

\begin{figure}[t]
\centering
\includegraphics[width=0.95\linewidth]{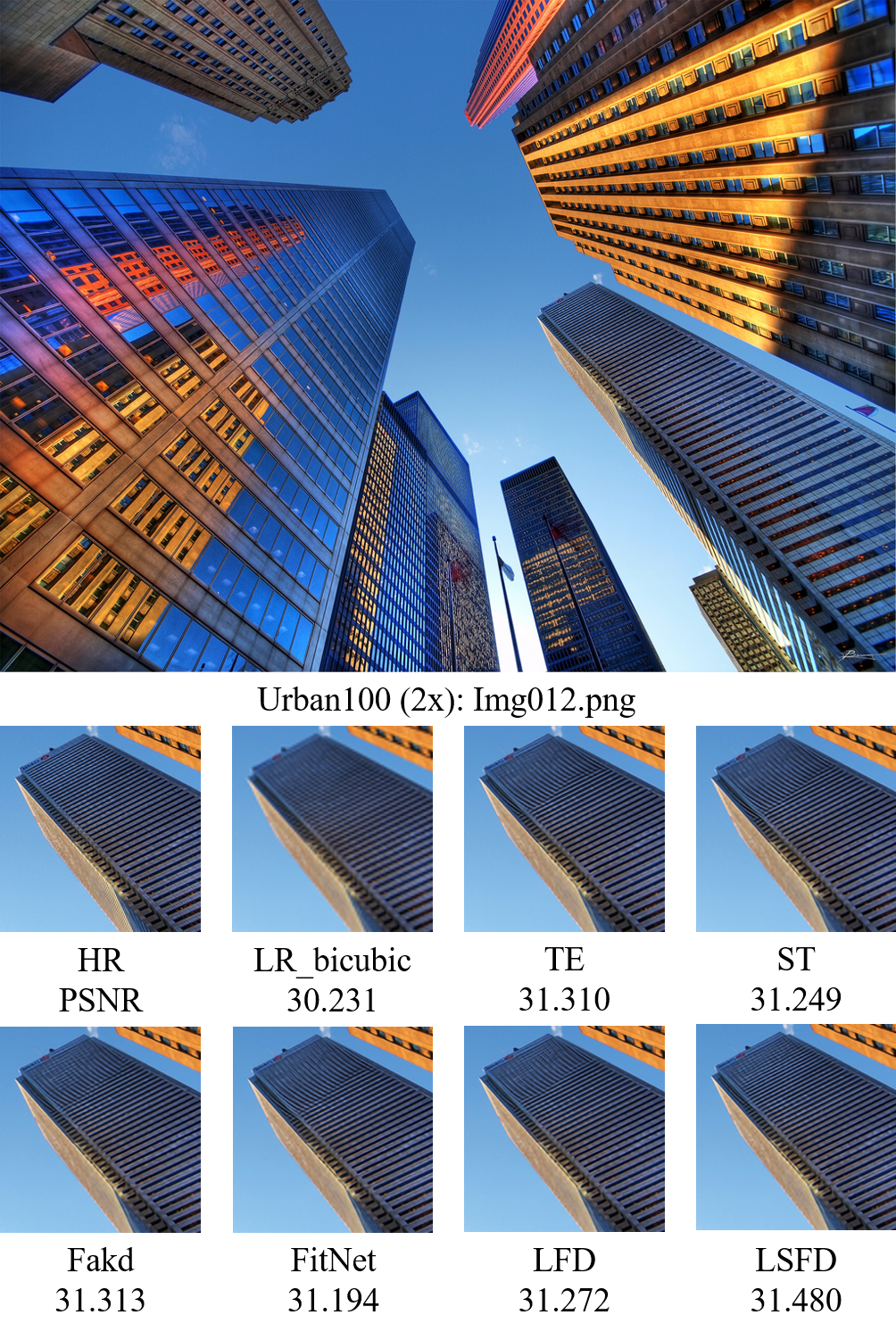}\\
\vspace{-4mm}
\caption{Visual comparison of different training methods on reconstructed images ($\times$2). }
\label{fig:x2}
\end{figure}

\begin{figure}[t]
\centering
\includegraphics[width=0.95\linewidth]{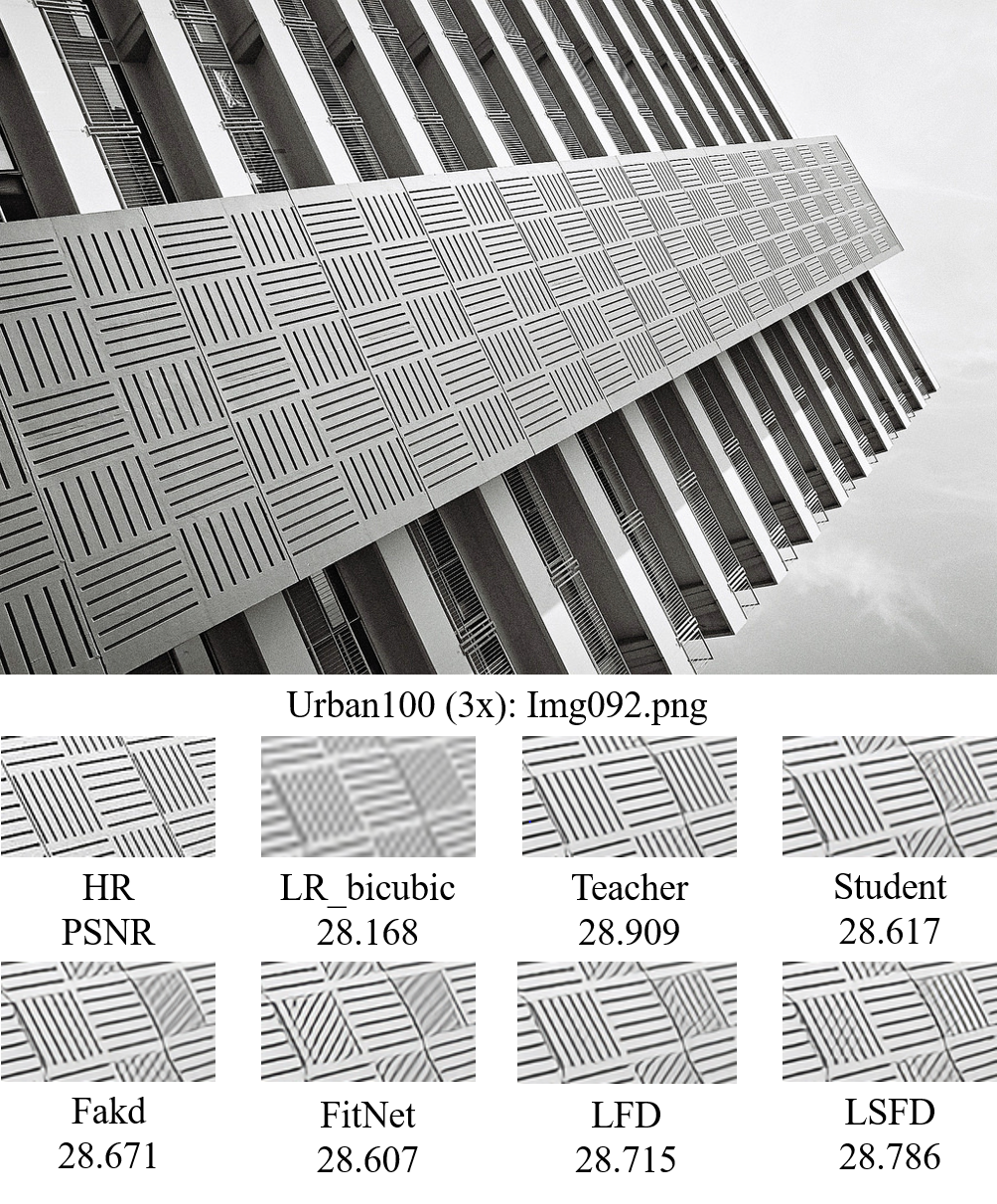}\\
\vspace{-4mm}
\caption{Visual comparison of different training methods on reconstructed images ($\times$3).}
\label{fig:x3}
\end{figure}

\begin{figure}[t]
\centering
\includegraphics[width=0.95\linewidth]{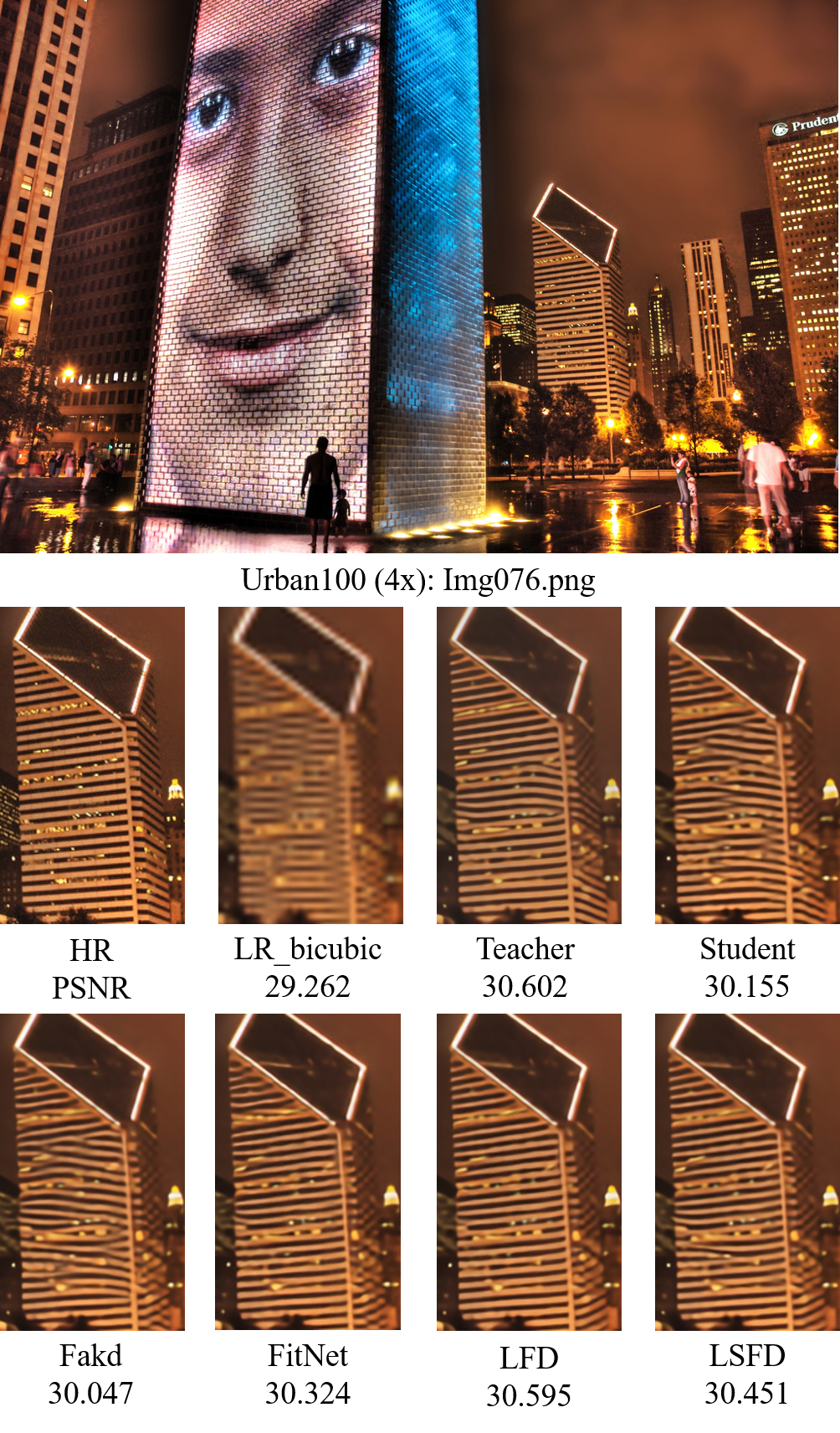}\\
\vspace{-4mm}
\caption{Visual comparison of different training methods on reconstructed images ($\times$4).}
\label{fig:x4}
\end{figure}

Detailed visualizations of the difficult Urban100 examples for different RCAN networks are provided in Figure \ref{fig:x2}, \ref{fig:x3} and \ref{fig:x4}, with each different images and upscale ratios. The images chosen are frequently used samples in other SISR papers. The `LR\_bicubic' image is attained by simply applying bicubic upsampling to the $I_{LR}$.
To compare the reconstruction power of the model with the PSNR scores directly, PSNR scores are calculated only using the cropped region for detailed comparison. 
As can be seen in the low-resolution images, {the} aliasing problem arises that distorts the original patterns, which mainly causes the performance degradation. 
For such images, the performance of the network is determined by how well it recovers the pattern of the HR using information from the surrounding patterns \cite{Gu_2021_CVPR}.
We can clearly see that distilled models produce better results. Especially, for Fig. \ref{fig:x2}, LSFD reconstructs better than the teacher network, which is consistent with the quantitative PSNR results that models trained {by} our method especially perform well on the Urban100 dataset.



\section{Discussion}
\noindent\textbf{Effective usage of receptive fields:} 
Various feature distillation methods exist that can be adapted to SISR, but many of them are shown to be ineffective in the paper of FAKD \cite{he2020FAKD}.
Since our method enables more expansive feature map areas to be involved for each distilled feature position, the student network is forced to focus on several surrounding positions in the feature map, which results in the effective usage of receptive fields, and this effect is visualized through the gradient map of the LAM.
In Figure \ref{fig:LAM}, we can infer that in the `Area of Contribution' of LSFD, our LSFD model tries to use \textbf{successfully generated area of itself as a reference} for generating the interested area.

\noindent\textbf{Limitations:} 
Although our model improves the performance of the student network a lot, its limitation is that features at different spatial locations that are far away cannot affect each other, even though they may have relations. 
For this property, although LSFD notably outperforms on anti-aliasing, as we verified on Urban100 experiments, they perform not as well on other datasets.
We think that it would be beneficial if the model could simultaneously attend to the more important feature map areas and simultaneously distill features based on the non-local operation. We tried some experiments that use attention to the input feature map of FAKD using the SFD matrix, but it led to performance degradation.

\section{Conclusion}
This paper described the importance of adapting KD to single image super-resolution (SISR) and proposed Local-Selective Feature Distillation (LSFD) that successfully applies the regressor-based feature distillation on the area of SISR. Our LFD showed the importance of utilizing a wider feature range in the receptive field point of view, and by additionally applying SFD, LSFD demonstrates the impact of our selective attention to more important feature areas both quantitatively and qualitatively. Our method suggests a promising viewpoint for future works adapting feature distillation to the task of super-resolution.
\clearpage

{\small
\bibliographystyle{ieee_fullname}
\bibliography{egbib}
}

\end{document}